\documentclass[
	aps,
	prx,
	floatfix,
	amsmath,amssymb,
	twocolumn,
	reprint
	superscriptaddress
]{revtex4-2}
\usepackage{graphicx}
\usepackage{dcolumn}
\usepackage{bm}
\usepackage[hidelinks]{hyperref}
\hypersetup{
  colorlinks = true, 
  urlcolor   = blue, 
  linkcolor  = blue, 
  citecolor  = blue  
}

\providecommand{\ignore}[1]{}

\begin{document}

\title{Spurious Antenna Modes of the Transmon Qubit}

\author{O. Rafferty}
\author{S. Patel}
\author{C. H. Liu}
\author{S. Abdullah}
\author{C. D. Wilen}
\author{D. C. Harrison}
\author{R. McDermott}
\email[]{rfmcdermott@wisc.edu}
\affiliation{Department of Physics, University of Wisconsin-Madison, Madison, Wisconsin 53706, USA}

\date{\today}

\begin{abstract}
Superconducting qubits are resonant absorbers of pair-breaking radiation. The metal pads that form the qubit capacitance support standing wave modes at frequencies of order 100~GHz; these modes are strongly coupled to free-space impedance through their electric dipole moment. While the antenna mode of the 3D transmon qubit is easily seen to be a resonant dipole, other 2D qubit types can be understood as the aperture duals of wire loop antennas or folded dipoles. For typical Josephson junction parameters, the junction provides a reasonable conjugate match to the fundamental antenna mode. We calculate the contribution to quasiparticle poisoning from resonant absorption of blackbody radiation. We extend our analysis to dissipation at the qubit frequency, where radiative losses provide an ultimate limit to qubit energy relaxation time. A clear understanding of the spurious antenna modes of qubits will allow designs that are insensitive to pair-breaking radiation and that display reduced radiative losses at the qubit frequency.
\end{abstract}

\maketitle

\section{Introduction\label{sec:level1}}

Superconducting qubits are implemented as low-loss, nonlinear microwave modes \cite{kjaergaard20}. The mode frequency is typically in the range 3-10~GHz; electromagnetic coupling of the qubit to the environment is carefully controlled in order to minimize decoherence. Environmental fluctuations at the qubit frequency induce relaxation, while low-frequency fluctuations induce qubit dephasing. Progress in understanding the electromagentic environment of the qubit has resulted in steady improvements in qubit fidelity over the last two decades.

Qubit devices are typically cooled to temperatures below 20~mK in order to suppress dissipation from quasiparticle excitations out of the superconducting ground state. At these temperatures, the equilibrium density of quasiparticles should be exponentially small. Nevertheless, researchers find quasiparticle density of order 1~$\mu$m$^{-3}$ \cite{martinis09, wang14, serniak18, serniak19}, tens of orders of magnitude larger than the thermal equilibrium quasiparticle density expected from theory. Nonequilibrium quasiparticles tunnel across the Josephson junctions, inducing both excitation and relaxation \cite{catelani11_prl}. Recent works have shown that gamma rays and cosmic rays can be a source of nonequilibrium quasiparticles \cite{vepsalainen20,wilen20}; while particle impacts give rise to damaging correlated errors due to phonon-mediated quasiparticle poisoning \cite{patel17}, the event rate is too low and the escape rate of athermal phonons from the chip too high to account for the background population of quasiparticles. Other studies point to pair-breaking radiation as a potentially significant source of quasiparticle generation \cite{serniak18, houzet19}. Researchers have demonstrated improvements by shielding devices from blackbody radiation and by introducing in-line filters to block pair-breaking photons from higher temperature stages \cite{barends11, corcoles11}. However, the mechanism by which photons couple to the qubit junction has not been understood.

In this letter, we show that transmon qubit structures \cite{koch07, houck08} act as resonant absorbers of pair-breaking radiation at frequencies far outside the qubit operating range, in the 10s of GHz to THz. While environmental fluctuations at these frequencies have no direct effect on qubit operation, they can break Cooper pairs, giving rise to quasiparticle poisoning that can degrade qubit coherence. The spurious antenna modes intrinsic to the qubit structure provide a perversely efficient route to channel quasiparticles to the qubit junction, where they do the most damage. We calculate the contribution to quasiparticle poisoning from the resonant absorption of blackbody radiation. In addition, our analysis provides a straightforward means to calculate the limit to qubit energy relaxation time imposed by radiation at the qubit frequency. We find that state-of-the-art qubit devices are approaching the limit to qubit $T_1$ imposed by radiative loss.

This paper is organized as follows. In Section \ref{sec:antenna}, we review the basics of antenna physics and describe Babinet's principle \cite{balanis16}, which allows us to map conventional 2D qubit geometries to dual wire antenna structures. In Section \ref{sec:modes}, we describe the fundamental resonant modes of specific qubit geometries that are pursued for quantum computing: the single-ended transmon with circular and rectangular island; the Xmon \cite{barends13}; the differential transmon \cite{chow12, place21}; and the 3D transmon \cite{paik11}. In Section \ref{sec:efficiency}, we consider the coupling of energy from the resonant antenna mode to the Josephson junction, and we show that for typical parameters the junction presents a reasonable congugate match to the radiation impedance of the antenna. In addition, we present detailed calculations of coupling efficiency and antenna noise bandwidth for specific qubit geometries. In Section \ref{sec:QP}, we present a simple analysis that allows us to quantify the contribution to quasiparticle generation at the junction from resonant absorption of broadband blackbody radiation. In Section \ref{sec:T1}, we quantify to the contribution to qubit dissipation from radiation at the qubit frequency, for which the qubit structure can be modeled as an electrically small antenna. Finally, in Section \ref{sec:conclude} we conclude with a brief discussion of implications of this analysis for other qubit types and of prospects for exploiting antenna coupling to pair-breaking photons to realize a new class of quantum sensors for dark matter detection and precision spectroscopy.

\section{Babinet's Principle: Wire/Aperture Duality \label{sec:antenna}}

Conducting structures that support time-varying currents and voltages will radiate efficiently at frequencies corresponding to standing wave resonances. The prototypical example is the half-wave dipole antenna; here, current is injected into a central feed point, which is a low-impedance current antinode. Current vanishes at the ends of the antenna arms; at the half-wave resonance, constructive interference leads to a standing wave. Radiation to the far field represents a real impedance, while energy stored in the near-field region represents an imaginary impedance. We will refer to this type of antenna, consisting of sparse conducting features that support time-varying currents that couple to the radiation field, as a \textit{wire antenna}.

While the 3D transmon \cite{paik11} is readily understood as a resonant half-wave dipole with fundamental resonance at a frequency of order 100~GHz, the dominant radiation mode of 2D planar qubits is not immediately obvious. The qubit is typically realized as a small superconducting island embedded in a near-continuous superconducting groundplane and connected to ground via one or more Josephson junctions. Due to symmetry, the lowest nonvanishing moment of the island+groundplane will be an electric quadrupole, which is expected to present a rather small radiation resistance. To understand the radiation mode of 2D transmon qubits, we use Babinet's principle \cite{balanis16}, which allows us to map from 2D qubit geometries to their wire antenna dual structures. We find that 2D qubits can be seen as the \textit{aperture antenna} duals of common wire antennas such as the folded half-wave dipole or the resonant loop antenna. With this insight, it is clear that 2D superconducting qubits, of the type currently explored for fault-tolerant arrays \cite{fowler12}, noisy intermediate-scale quantum processors \cite{preskill18}, or demonstrations of quantum supremacy \cite{arute19}, will radiate at a fundamental frequency where the wavelength is matched to the qubit perimeter.
\begin{figure}
\includegraphics[width=\columnwidth]{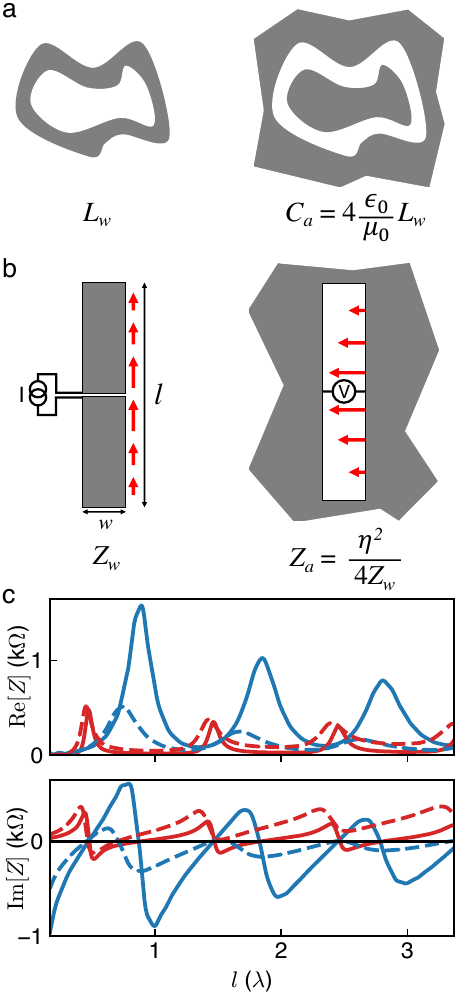}
\caption{\label{fig:Fig1} Wire/aperture duality. (a) The planar loop in wire space is dual to the planar capacitor embedded in a groundplane in aperture space. (b) Planar wire antenna and its aperture dual. The antennas have length $l$ and width $w$, both large compared to film thickness. Here, the direction and magnitude of $E$-fields at the fundamental modes are shown schematically by red arrows. (c) Simulated real and imaginary parts of the antenna impedance $Z$ of the wire (blue) and aperture (red) structures at aspect ratio $l/w\approx 200$ (solid) and $l/w\approx 20$ (dashed).}

\end{figure}

According to Babinet's principle, a wire antenna is dual to the aperture structure where conductor is replaced by empty space in an otherwise continuous conducting plane. If the wire antenna is fed by a current injected into a low-impedance current antinode, feed in the aperture dual is a voltage coupled to the high-impedance voltage antinode. The complex radiation impedances of the wire antenna $Z_w$ and its aperture dual $Z_a$ are related by the formula

\begin{equation}
    Z_a Z_w= \frac{\eta^2}{4},
\end{equation}
where $\eta = \sqrt{\mu_0/\epsilon_0} \approx 377 \, \Omega$ is the impedance of free space. As an example, the capacitance $C_a$ of a planar electrode embedded in a groundplane in aperture space is simply related to the inductance $L_w$ of the complementary loop in wire space: $C_a = 4(\epsilon_0/\mu_0)L_w$ (see Fig. 1a).

Here we are concerned with transmon qubit structures fabricated using thin-film processing techniques on planar substrates. As an example of the dual mapping from wire antenna to aperture antenna, we consider in Fig. \ref{fig:Fig1}b a wire antenna implemented using rectangular thin-film electrodes with length $l$ and width $w$, both assumed to be large compared to film thickness; this structure is the basis for the 3D transmon qubit \cite{paik11}, considered in detail below. The structure is dual to the aperture or slot antenna shown in the right panel of Fig. \ref{fig:Fig1}b. In mapping from the wire antenna to the aperture dual, a low-impedance point is mapped to high impedance and vice versa; in addition, the polarization of the radiated field is rotated by 90$^\circ$. However the standing wave mode and the dipole radiation pattern are the same. In Fig. \ref{fig:Fig1}c, we plot the real and imaginary parts of the radiation impedance of these two antennas calculated using a finite-element solver \cite{cst} for aspect ratio $l/w$~=~200 and 20. In the limit of large $l/w$, the wire antenna reduces to the ideal resonant half-wave dipole. For structures with lower aspect ratio, sharp resonant structure in $Z$ is smoothed out, and resonances are shifted to lower frequency. Peaks in the radiation resistance of the wire antenna correspond to low radiation resistance for the aperture dual, and vice versa.

\begin{figure}[t]
\includegraphics[width=\columnwidth]{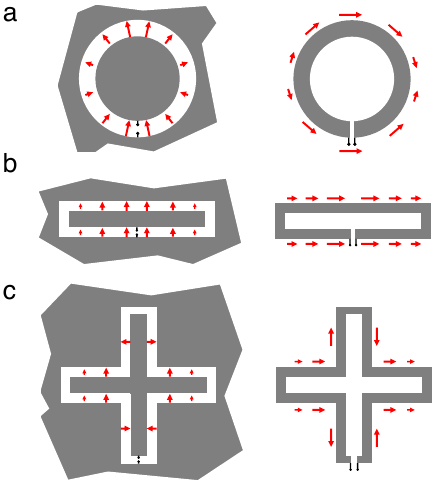}
\caption{\label{fig:Fig2} Typical 2D qubit structures and their wire antenna duals. (a) The transmon with circular island is dual to the resonant loop antenna. (b) The transmon with rectangular island is dual to the folded half-wave dipole antenna. (c) The Xmon qubit \cite{barends13} supports a resonant antenna mode corresponding to device perimeter equal to one full wavelength. For the structure shown here, the horizontal arms radiate like a folded dipole, while the vertical arms do not couple to the far field. In this figure, the direction and magnitude of $E$-fields at the fundamental modes are shown schematically by red arrows.
}
\end{figure}

\section{Resonant Antenna Modes of Typical Transmon Structures \label{sec:modes}}

In Fig. \ref{fig:Fig2}, we show simplified layouts of various 2D qubit geometries along with their wire duals. In the case of a circular transmon island, the structure is the dual of the conventional loop antenna, which involves a fundamental resonance when the loop circumference is matched to the wavelength. The transmon qubit with rectangular island is mapped to the resonant half-wave folded dipole antenna. Here, for the aperture (wire) structure, there are voltage (current) nodes at the ends of the antenna arms, so that $E$-fields interfere constructively across the arms of the structure. The Xmon structure of \cite{barends13} is mapped to a variant of the folded dipole antenna, involving a fundamental resonance where $E$-fields from the horizontal arms interfere constructively, while $E$-fields from the vertical arms interfere destructively.

\begin{figure}
\includegraphics[width=\columnwidth]{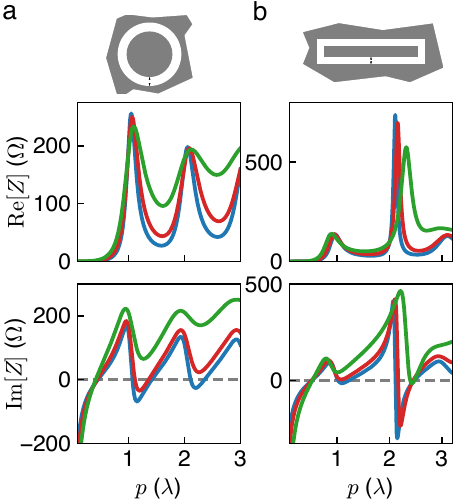}
\caption{\label{fig:Fig3} Antenna impedance of single-ended transmons. (a) Real and imaginary parts of the antenna impedance of circular transmons with aspect ratios $p/w$~=~100 (blue), 50 (red), and 20 (green), where $p$ is the perimeter at the center of the gap and $w$ is the width of the gap. (b) Real and imaginary parts of the antenna impedance of rectangular transmons with aspect ratios and trace colors as in (a). }
\end{figure}

In Fig. \ref{fig:Fig3}, we plot the real and imaginary parts of the antenna impedance for the circular and rectangular transmon geometries, for various device aspect ratios. The fundamental full-wave resonance corresponds to a peak in the real part of the antenna impedance. In the vicinity of this resonance, the antenna presents an inductive impedance with positive imaginary part. As we will show below, such a structure allows an excellent conjugate match to the impedance of a Josephson junction, providing an efficient means to couple pair-breaking radiation directly to the junction.

\begin{figure*}
\includegraphics[width=\textwidth]{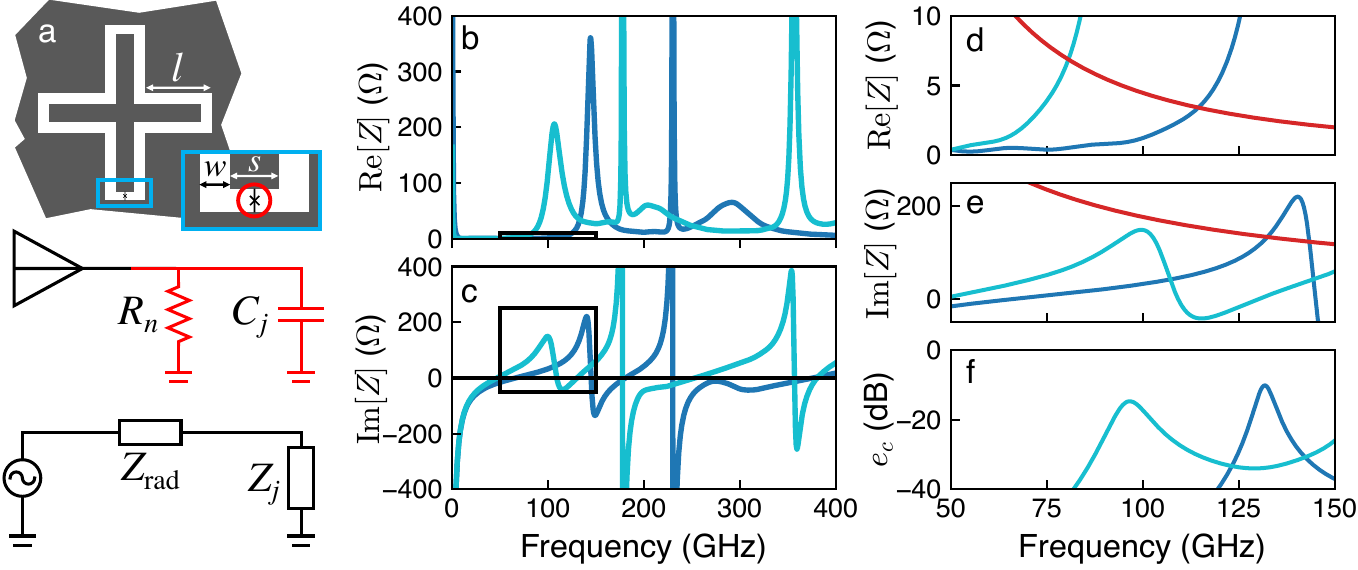}
\caption{\label{fig:xmon} Resonant antenna mode of the Xmon qubit. (a) Geometry of the Xmon qubits described in \cite{barends13}. The center trace width is $s$; gap to the groundplane is $w$; and length of a single arm is $l$. The junction is modeled as a real tunnel resistance $R_n$ in parallel with the junction self-capacitance $C_j$. Real (b) and imaginary (c) parts of Xmon antenna impedance for devices with $l$~=~130~$\mu$m, $s$~=~8~$\mu$m, and $w$~=~4~$\mu$m (blue) and  $l$~=~165~$\mu$m, $s$~=~24~$\mu$m, and $w$~=~24~$\mu$m (cyan). Real (d) and imaginary (e) parts of $Z_{\rm rad}$ and $Z_j^*$ (red) in the vicinity of the fundamental antenna resonance. Here we take $R_n$ = 7~k$\Omega$ and $C_j$~=~ 9~fF. (f) Frequency-dependent coupling efficiency $e_c$ for the two Xmon devices. For the larger device, the coupling efficiency peaks at 97~GHz with antenna noise bandwidth $\Delta f_N$~=~1.8~GHz. For the smaller device, coupling efficiency peaks at 130~GHz with noise bandwidth $\Delta f_N$~=~2.5~GHz.}
\end{figure*}

We note that the wire/aperture duality extends to the dielectric or magnetic medium in which the antenna is embedded: relative permittivity in one space is mapped to relative permeability in the complementary dual space. That is, for relative permittivity and permeability $\epsilon_r, \mu_r$ in the aperture antenna space and relative permittivity and permeability $\epsilon_r', \mu_r'$ in the dual wire antenna space, we have $\epsilon_r'=\mu_r, \, \mu_r'=\epsilon_r$. Here we are concerned with aperture antenna structures fabricated on dielectric substrates with relative permittivity $\epsilon_r$. In the limit of an infinitely thick substrate, this is equivalent to the same antenna structure embedded in a uniform dielectric medium with effective permittivity $\epsilon_{\mathrm{eff}} = \left(1+\epsilon_r\right)/2.$ This structure is then mapped to a wire antenna embedded in a magnetic medium with permeability $\mu_r' = \epsilon_{\rm{eff}}$. In the following analysis of specific 2D qubit geometries, we consider an infinitely thick substrate with relative permittivity $\epsilon_r = 11$, a reasonable match to either silicon ($\epsilon_r$ = 11.7) or sapphire (anisotropic; $\epsilon_r$ = 8.9-11.1). For dielectric substrates of finite thickness, standing waves in the substrate can give rise to additional resonant structure in the radiation impedance of the antenna. In general, accurate modeling will require full numerical simulation that takes into account the substrate dimensions, including placement of the qubit structure on the substrate chip.

\section{Coupling Efficiency to the Josephson Junction \label{sec:efficiency}}

We are concerned with absorption of radiation at a frequency above the superconducting gap; in this case, we can model the junction as a resistance equal to the tunnel resistance $R_n$ (of order 5-10~k$\Omega$) shunted by the junction self-capacitance $C_j$ (typically 1-10~fF). The tunnel resistance is related to the critical current $I_0$ via the Ambegaokar-Baratoff relation \cite{tinkham96}:
\begin{equation}
    R_n = \frac{\pi \Delta}{2eI_0}.
\end{equation}
For aluminum, we have $2\Delta/e = 380~\mu$V and $I_0 R_n \approx$~300~$\mu$V. We can map the junction resistance and self capacitance to a complex impedance $Z_j$ consisting of a (frequency-dependent) real resistance in series with a negative imaginary impedance:

\begin{equation}
    Z_j = \frac{1 - j \omega \tau}{1 + \omega^2 \tau^2} R_n,
\end{equation}
where $\tau \equiv R_nC_j$. As a result, the junction presents a capacitive load to the antenna, with radiation impedance $Z_{\rm rad}$. Optimal power transfer is achieved when the conjugate matching condition is satisfied: $Z_{\rm rad} = Z_j^*$. In the case of mismatch between the junction and the antenna, the power transferred to the junction is reduced compared to the maximum available power. We define coupling efficiency $e_c$ as follows:
\begin{equation}
    e_c = 1-\left|\Gamma\right|^2,
\end{equation}
where
\begin{equation}
    \Gamma = \frac{Z_{\mathrm{rad}} - Z_j^*}{Z_{\mathrm{rad}} + Z_j}.
\end{equation}

We define an antenna noise bandwidth $\Delta f_N$ by integrating $e_c$ over frequency:
\begin{equation}
    \Delta f_N = \int e_c \, df.
    \label{eq:bw}
\end{equation}
As we will see in Section \ref{sec:QP} below, this noise bandwidth determines the integrated power of pair-breaking radiation coupled to the junction.

\subsection{Xmon Qubit \label{subsec:Xmon}}

In Fig. \ref{fig:xmon} we plot $Z_\mathrm{rad}$ and $Z_j^*$ for two of the Xmon qubit devices described in \cite{barends13}. The device dimensions are taken from the manuscript; the (compound) junction tunnel resistance $R_n$ = 7~k$\Omega$ is inferred from the reported mode frequency and anharmonicity, and the junction capacitance $C_j$ = 9 fF is taken from the reported area of the junctions, using a specific capacitance of 75~fF/$\mu$m$^2$. We assume a semi-infinite substrate with relative permittivity $\epsilon_r=11$, corresponding to $\epsilon_{\rm eff} = 6$. For the device with narrow traces and narrow gap, we find reasonable power match to the junction at a frequency of 130~GHz and a noise bandwidth $\Delta f_N$~=~1.8~GHz. For the device with wider trace width and wider gap, the resonance is pulled downward to around 97~GHz and we find $\Delta f_N$~=~2.5~GHz.

\subsection{Differential Qubits \label{subsec:diff}}

\begin{figure}
\includegraphics[width=\columnwidth]{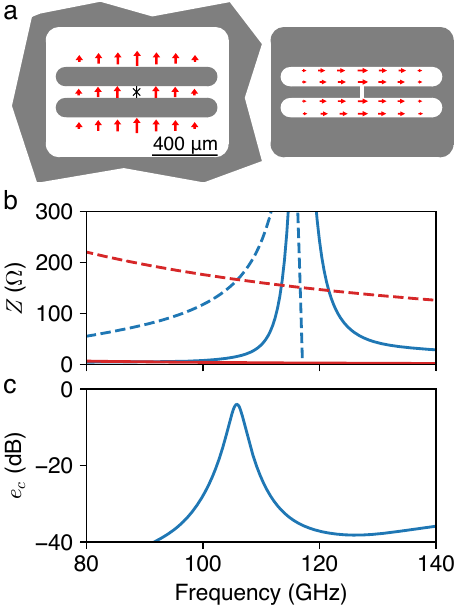}
\caption{\label{fig:diff} Resonant antenna mode of the differential transmon. (a) Simplified layout of the differential transmon and its wire antenna dual. The fundamental resonance corresponds to round trip around one island equal to a full wavelength. Here the direction and magnitude of $E$-fields at the fundamental resonance are shown schematically by red arrows. (b) Real (solid blue) and imaginary (dashed blue) parts of antenna impedance $Z_{\rm rad}$ of the differential qubit structure of \cite{place21}, along with real (solid red) and imaginary (dashed red) parts of the conjugate junction impedance $Z_j^*$ for $R_n$~=~7~k$\Omega$ and $C_j$~=~9~fF. (c) Frequency-dependent coupling efficiency $e_c$ calculated from (b).
}
\end{figure}

A number of groups use differential qubits, where the qubit capacitance is formed from two superconducting islands with no galvanic connection to the circuit groundplane \cite{chow12}. It has been argued that this symmetric construction provides additional protection against environmental fluctuations \cite{steffen09}. The resonant radiation mode of these structures can be understood simply from the wire antenna dual. In Fig. \ref{fig:diff}a, we show schematic diagrams of a differential qubit along with its wire dual. The wire dual can be viewed as a ``doubled" folded half-wave dipole antenna, with a single feed branch and two parallel return branches. The radiation pattern and radiation resistance are expected to be identical to those for the conventional folded dipole. We have calculated the radiation resistance $Z_{\rm rad}$ for the differential qubit shown in Fig. \ref{fig:diff}a, which is a slight simplification of the qubit geometry described in \cite{place21}; results are shown in Fig. \ref{fig:diff}b, along with $Z_j^*$ corresponding to a junction with $R_n$~=~7~k$\Omega$ and $C_j$~=~9~fF. The folded dipole resonance of the structure occurs at 110~GHz, and the device presents an excellent conjugate match to the junction. In Fig. \ref{fig:diff}c, we plot the antenna coupling efficiency for this qubit in the vicinity of the fundamental resonance. We find an antenna noise bandwidth $\Delta f_N$~=~2.8~GHz, comparable to the single-ended Xmon qubit.

At low frequencies, these structures are dual to electrically small magnetic quadrupoles. Symmetry-based suppression of the dipole moment will provide protection against dissipation from radiation. This case is discussed in more detail in Section \ref{sec:T1} below.

\subsection{3D Transmon \label{subsec:3D}}
\begin{figure}
\includegraphics[width=\columnwidth]{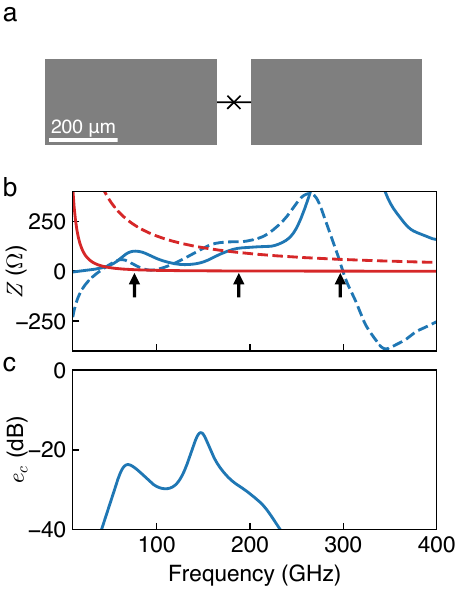}
\caption{\label{fig:3D} Resonant dipole antenna mode of the 3D transmon. (a) Device geometry (from \cite{paik11}). (b) Real (solid blue) and imaginary (dashed blue) parts of antenna impedance $Z_{\rm rad}$ of the 3D qubit, along with real (solid red) and imaginary (dashed red) parts of the conjugate junction impedance $Z_j^*$ for $R_n$~=~7~k$\Omega$ and $C_j$~=~9~fF. Arrows indicate the $\lambda/2$, $3\lambda/2$, and $5\lambda/2$ dipole resonance frequencies calculated from the dimensions of the structure. (c) Frequency-dependent coupling efficiency $e_c$ calculated from (b).
}
\end{figure}

The dominant radiation mode of the 3D transmon is that of a resonant half-wave dipole. In Fig. \ref{fig:3D}a we reproduce the geometry of the qubit described in \cite{paik11}, and in Fig. \ref{fig:3D}b we plot $Z_{\rm rad}$ and $Z_j^*$ for $R_n$~=~7~k$\Omega$, $C_j$~=~9~fF. Here we have taken $\epsilon_{\rm eff}$~=~6, which would correspond to an infinitely thick substrate with $\epsilon_r$~=~11. Given the large dimensions of the qubit capacitor pads, this is clearly an oversimplification; more detailed modeling would be required for an accurate extraction of $\epsilon_{\rm eff}$. Within our simplified model, we find a reasonable conjugate match to the junction at a frequency of 150~GHz, with antenna noise bandwidth $\Delta f_N$~=~4.4~GHz.

\section{Implications for Quasiparticle Poisoning \label{sec:QP}}

Thus, typical transmon structures act as highly efficient resonant absorbers of pair-breaking radiation. The coupling of pair-breaking photons to the spurious antenna mode is particularly damaging for the qubit, as the generated quasiparticles are created directly at the junction. The photon-assisted quasiparticle generation mechanism can induce both upward and downward qubit transitions \cite{houzet19}. Moreover, due to spatial inhomogeneities in the superconducting gap, the generated quasiparticles might remain localized near the junction, where they could tunnel back and forth repeatedly, giving rise to additional qubit transitions.

To connect the power spectral density of pair-breaking photons to a rate of quasiparticle generation at the qubit junction, we assume coupling of the qubit antenna to a single polarization and a single mode of the blackbody radiation field at temperature $T$. The blackbody power spectral density is given by
\begin{equation}
    S (f,T) = \frac{h f}{e^{hf/k_BT}-1}.
\end{equation}
The total power coupled to the qubit junction is determined by integrating over the frequency-dependent coupling efficiency $e_c(f)$:
\begin{equation}
\begin{aligned}
    P &=& \int S(f, T) \, e_c(f) \, df \\
      &\approx& \frac{h f_0}{e^{hf_0/k_BT}-1} \, \Delta f_N,
\end{aligned}
\end{equation}
where $f_0$ is the frequency of maximum power transfer between the qubit antenna structure and the  junction, and where $\Delta f_N$ is defined in Eq. \ref{eq:bw} above. The rate $\Gamma_{\rm pa}$ of photon-assisted quasiparticle poisoning events at the junction is therefore given by
\begin{equation}
    \Gamma_{\rm pa} = \frac{\Delta f_N}{e^{hf_0/k_BT}-1}.
\end{equation}
We take the larger Xmon qubit of Section \ref{subsec:Xmon} as an example. For $f_0$~=~97~GHz and $\Delta f_N$~=~1.8~GHz, we find that a rate $\Gamma_{\rm pa}$~=~300~Hz of quasiparticle poisoning events (taken, e.g., from \cite{christensen19}) corresponds to an effective blackbody temperature $T$~=~300~mK.

All transmon qubits analyzed here involve a resonant antenna mode at a frequency of order 100~GHz; this frequency is set by the overall scale of the device. As devices typically target a rather narrow range of qubit charging energy, it is understandable that the fundamental resonant antenna modes occupy a similarly small range of frequency. Moreover, typical junctions are reasonably well matched to the radiation impedance of the antenna mode, yielding noise bandwidth in the range 2-5~GHz. The consistency of antenna mode frequency and bandwidth across different qubit types leads us to conclude that coupling to blackbody radiation with effective temperature of order 300~mK is a likely explanation for the excess quasiparticles observed in superconducting qubit experiments.

\section{Dissipation at the Qubit Frequency \label{sec:T1}}
The mapping of an aperture qubit to its wire antenna dual provides a particularly simple means to calculate the contribution to qubit dissipation from radiation. Qubit $T_1$ time is calculated from the real part of the admittance shunting the junction. If we neglect sources of loss other than radiation to the environment, we have
\begin{equation}
    T_1 = \frac{C}{\mathrm{Re}[Y_a(\omega_{01})]},
\end{equation}
where $C$ is the qubit capacitance, $Y_a$ is the radiation admittance of the aperture antenna formed by the qubit island and groundplane, and $\omega_{01}$ is the qubit transition frequency. From Babinet's principle, we have $Y_a = 4Z_w/\eta^2$, where $Z_w$ is the radiation impedance of the wire antenna that is dual to the qubit structure. As a result, we have
\begin{equation}
    T_1 = \frac{\eta^2}{4}\frac{C}{\mathrm{Re}[Z_w(\omega_{01})]}.
\end{equation}
Thus, knowledge of the radiation resistance of the wire antenna that is dual to the qubit structure allows straightforward calculation of the radiation limit to qubit $T_1$.

We consider the simple case of a transmon with circular island embedded in a circular cavity in a continuous groundplane; we assume an effective permittivity $\epsilon_{\rm{eff}}$. The structure is the aperture dual of a wire loop antenna embedded in a magnetic medium with relative permeability $\mu_r' = \epsilon_{\rm{eff}}$. We first take the limit of an infinitesimal wire width, for which radiation resistance can be calculated analytically:
\begin{equation}
    \mathrm{Re}[Z_w] = \frac{8}{3}\pi ^5 \epsilon_{\rm eff}^{1/2} \eta \left(\frac{r}{\lambda}\right)^4.
\end{equation}
We find a limit to qubit $T_1$ given by
\begin{equation}
    T_1 = \frac{3}{2\pi} \epsilon_{\rm eff}^{-5/2} \eta C \left(\frac{c}{\omega_{01}r}\right)^4,
    \label{eq:T1}
\end{equation}
where $c$ is the speed of light in vacuum.

For a circular transmon island with radius $r_i$ embedded in a circular cavity in the groundplane with finite gap width $w$, we calculate $\mathrm{Re}[Z_w]$ numerically using the method outlined in \cite{sendelbach08_notes}. For aspect ratio $w/r_i \lesssim 0.2$, the radiation resistance of the structure is well approximated by
\begin{equation}
    \mathrm{Re}[Z_{w}] =  \frac{8}{3}\pi ^5 \epsilon_{\mathrm {eff}}^{1/2} \eta \left(\frac{r_i}{\lambda}\right)^4 \left(1 + 2.1 \frac{w}{r_i}\right).
\end{equation}
For typical qubit parameters, the correction to Re$[Z_w]$ due to finite gap width will be of order 10-30\%.

For arbitrary qubit island geometry, we employ the same analysis, with the replacement $\pi r_i^2 \rightarrow A$, where $A$ is the area of the qubit island.

Returning to Eq. \ref{eq:T1}, if we take $C$~=~100~fF, $\omega_{01}/2\pi$~=~5~GHz, and $r$~=~100~$\mu$m, we find a radiative limit to qubit $T_1$ of 1.5~ms for $\epsilon_{\rm eff}$~=~1, and of 17~$\mu$s for $\epsilon_{\rm eff}$~=~6. The latter choice of $\epsilon_{\rm eff}$ corresponds to an infinitely thick substrate with relative permittivity $\epsilon_r$~=~11. For a device fabricated on a substrate with thickness of order several hundred microns, considerably less than the wavelength at the qubit frequency, we expect that the effective permittivity will suppressed from the value $\epsilon_{\rm eff} = (\epsilon_r + 1)/2$. More detailed modeling of finite substrate effects will be needed to achieve a quantitative understanding of the radiative limit to qubit $T_1$, which scales with effective permittivity as $\epsilon_{\rm eff}^{-5/2}$.

In the case of the differential qubit \cite{place21}, radiative losses are suppressed by symmetrization of the structure: the wire dual can be viewed as a magnetic quadrupole. For practical devices, however, the symmetry of the structure will be broken, and radiative losses will be dominated by the magnetic dipole moment associated with the area of unsymmetrized structures.

For the 3D transmon qubit, the radiation limit to $T_1$ is very severe for a device looking out at free space, as radiation from an electric dipole scales as $\left(r/\lambda\right)^2$. However, encapsulation of the qubit chip in a conducting cavity provides a Purcell suppression of the environmental density of states that protects the qubit from radiative loss \cite{houck08}.

\section{Conclusion \label{sec:conclude}}

In conclusion, we have described the resonant coupling of superconducting qubit structures to pair-breaking photons via a spurious antenna mode. While the analysis presented here focuses on the transmon qubit, it can readily be extended to other qubit types, including fluxonium \cite{manucharyan09} and the capacitively shunted flux qubit \cite{yan16}. The same considerations can be applied to microwave kinetic inductance detectors based on disordered films of granular aluminum \cite{henriques19} or nitride-based superconductors \cite{barends10}, for which the high normal state resistance of the metal film will provide a good power match to the radiation impedance of the antenna. It is quite possible that approaches to topologically protected qubits based on Majorana particles \cite{lutchyn10}, for which quasiparticle poisoning is fatal, will be affected by the antenna coupling described here.

We extended our analysis to the case of radiation at the qubit frequency, for which the qubit can be modeled as an electrically small antenna. Our modeling did not attempt to capture the impact of finite substrate thickness on the effective permittivity of the qubit medium, which will have a strong influence on radiative loss. However, even in the most optimistic scenario corresponding to $\epsilon_{\rm eff}$~=~1, we find a limit to qubit $T_1$ from radiation of order 1~ms. While the desire to protect against loss due to dielectric defects at interfaces motivates  a push towards larger device scales \cite{martinis05}, the need to limit radiative loss will provide an ultimate constraint on the size of qubit devices.

A detailed understanding of antenna coupling to pair-breaking photons will allow the engineering of optimized detectors of photons in the 10s of GHz to THz range for the purposes of narrowband spectroscopy of the cosmic microwave background or for detection of dark matter axions \cite{braine20}. Antenna coupling to the radiation field will allow controlled transduction of photons to quasiparticles at the junction. These quasiparticles could be detected directly, or superconducting gap engineering \cite{aumentado04} could be used to guide the generated quasiparticles to a separate qubit structure that would detect the change of quasiparticle parity with high fidelity \cite{riste13, serniak18, christensen19}.

\smallskip

We thank John M. Martinis for stimulating discussions. Work supported by the Fermi National Accelerator Laboratory, managed and operated by Fermi Research Alliance, LLC under Contract No. DE-AC02-07CH11359 with the U.S. Department of Energy, through the Office of High Energy Physics QuantISED program.


%

\end{document}